\documentclass[conference]{IEEEtran}
\ifx\pdfoutput\undefined
\bibliography{mybib}
\usepackage{graphicx}
\else
\usepackage[pdftex]{graphicx}
\fi
\usepackage{url}
\usepackage{array}
\usepackage{stfloats}
\usepackage{amssymb}
\usepackage{amsmath}
\usepackage{amsthm}
\usepackage{cite}
\usepackage{enumerate}
\usepackage{epstopdf}
\usepackage{color}

\DeclareGraphicsExtensions{.pdf,.png,.jpg}
\begin{document}
 \title{ Energy Efficient Symbol-Level Precoding  in  Multiuser MISO Channels}
  
 
 \author{%
 Maha Alodeh, \quad Symeon Chatzinotas\quad Bj\"{o}rn Ottersten\\

Interdisciplinary Centre for Security, Reliability and Trust (SnT),
University of Luxembourg\\

e-mail:\{maha.alodeh, symeon.chatzinotas, and bjorn.ottersten\}@uni.lu\thanks{This work is supported by Fond National de la Recherche Luxembourg (FNR)
projects, project Smart Resource Allocation for Satellite Cognitive Radio (SRAT-SCR)  ID:4919957 and Spectrum Management and Interference Mitigation in Cognitive Radio Satellite Networks SeMiGod, SATellite SEnsor NeTworks for spectrum monitoring (SATSENT).}}
 
 \maketitle
 \IEEEpeerreviewmaketitle
 \begin{abstract}
\boldmath This paper investigates the idea of exploiting interference among the simultaneous multiuser transmissions in the downlink of multiple
antennas systems. Using symbol level precoding, a new approach towards addressing the multiuser interference is discussed through jointly utilizing the channel state information (CSI) and data information (DI). In this direction, the interference among the data streams
is transformed under certain conditions to useful signal that can improve the signal to interference noise ratio (SINR) of the downlink transmissions. In this context, new constructive interference precoding techniques that tackle the transmit power minimization (min power) with
individual SINR constraints at each user's receivers are proposed. Furthermore, we investigate the CI precoding design under the assumption that the received MPSK symbol can reside in a relaxed region in order to be correctly detected. 
 Finally, extensive numerical results
show that the proposed schemes outperform other state of the art techniques.
\end{abstract}

\vspace{-0.1cm}
\section{Introduction}
Interference
is one of the crucial and limiting factors in wireless networks. The idea of utilizing the time and frequency resources has been proposed in the literature to allow different users to share the wireless medium without inducing harmful interference. The
concept of exploiting the users' spatial separation has been a fertile research domain for more than one
decade\cite{roy}. This can be implemented by adding multiple antennas at one or both
communication sides. Multiantenna transceivers empower the communication systems with more degrees
of freedom that can boost the performance if the multiuser interference is mitigated
properly. Exploiting the space dimension, to serve different users simultaneously
in the same time slot and the same frequency band through spatial division multiplexing (SDMA), has been investigated in \cite{roy}-\cite{haardt}. 

The main idea is to exploit the spatial multiuser interference by redesigning the interference to be constructively detected at the users' terminal. The scheme is different from the conventional techniques \cite{haardt}, which they decouple the multiuser transmission to reduce the interference power received at each terminal.  In \cite{Christos-1}, the interference in the scenario of BPSK and QPSK is classified into types: constructive and destructive. Based on this classification, a selective channel inversion scheme is proposed to eliminate the destructive interference while retaining the constructive one to be received at the users' terminal. A more elaborated scheme is proposed in \cite{Christos}, which  rotates the destructive interference to be received as useful signal with the constructive one. These schemes outperform the conventional precodings \cite{haardt} and show considerable gains. However, the anticipated
gains come at the expense of additional complexity at the system design level. Assuming
that
the channel coherence time is $\tau_{c}$, and the symbol period is $\tau_s$, with $\tau_c\gg\tau_s$ for slow fading
channels, the user precoder has to be recalculated with a frequency of $\frac{1}{\tau_c}$
in comparison with the symbol based precoder $\frac{1}{\min(\tau_c,\tau_s)}=\frac{1}{\tau_s}$. Therefore, faster precoder calculation and switching is needed in
the symbol-level precoding which can be translated to more expensive hardware.\smallskip

In this direction, we propose a symbol based precoding to exploit the interference by establishing the connection between the constructive interference precoding and multicast\cite{maha}-\cite{maha_TSP}. However, the precoding techniques design the received to be detected at the exact constellation point. 
In this work, we aim at optimizing the constructive interference among the spatial streams while we allow flexible precoding design.  We exploit the fact that the received symbol should lie in the correct detection region but not necessarily at the exact constellation point. This enables flexibility at the precoding design level in comparison with \cite{maha}-\cite{maha_TSP}, where they design the precoding to make the symbols detectable at the exact constellation (i.e. if we exclude the noise effect).   \smallskip
\smallskip 

\textbf{Notation}:  We use boldface upper and lower case letters for
 matrices and column vectors, respectively. $(\cdot)^H$, $(\cdot)^*$
 stand for Hermitian transpose and conjugate of $(\cdot)$. $\mathbb{E}(\cdot)$ and $\|\cdot\|$ denote the statistical expectation and the Euclidean norm,  $\mathbf{A}\succeq \mathbf{0}$ is used to indicate the positive
semidefinite matrix. $\angle(\cdot)$, $|\cdot|$ are the angle and magnitude  of $(\cdot)$ respectively. $\mathcal{R}(\cdot)$, $\mathcal{I}(\cdot)$
 are the real and the imaginary part of $(\cdot)$. Finally, the vector of
 all zeros with length of $K$ is defined as $\mathbf{0}^{K\times 1}$.
\vspace{-0.15cm} 
\section{System and Signal Models}
\label{system}
We consider a single-cell multiple-antenna downlink scenario,
where a single BS is equipped with $M$
transmit antennas that serves $K$ user terminals,
each one of them equipped with a single receiving antenna. The adopted
modulation technique is M-PSK.
We assume a quasi static block fading channel $\mathbf{h}_j\in\mathbb{C}^{1\times
M}$ between
the BS antennas and the $j^{th}$ user, where the received signal at
j$^{th}$ user is
written as
\vspace{-0.3cm}
\begin{eqnarray}
y_j[n]&=&\mathbf{h}_j\mathbf{x}[n]+z_j[n].
\end{eqnarray} $\mathbf{x}[n]\in\mathbb{C}^{M\times 1}$ is the transmitted signal vector from the multiple antennas
transmitter and  $z_j$ denotes the noise at $j^{th}$ receiver, which is assumed i.d.d  complex Gaussian distributed variable $\mathcal{CN}(0,1)$. A compact formulation
of the received signal at all users' receivers can be written as
\vspace{-0.1cm}
\begin{eqnarray}
\mathbf{y}[n]&=&\mathbf{H}\mathbf{x}[n]+\mathbf{z}[n].
\end{eqnarray}
Let $\mathbf{x}[n]$ be written as $\mathbf{x}[n]=\sum^K_{j=1}\mathbf{w}_j[n]d_j[n]$,
where $\mathbf{w}_j$ is the $\mathbb{C}^{M\times
1}$ unit power precoding vector for the user $j$. The received signal at $j^{th}$
user ${y}_j$ in $n^{th}$ symbol period is given by
\begin{eqnarray}
\label{rx_o}
\hspace{-1cm}{y}_j[n]=\sqrt{p_j[n]}\mathbf{h}_j\mathbf{w}_j[n] d_j[n]+\displaystyle\sum_{k\neq j}\sqrt{p_k[n]}\mathbf{h}_j\mathbf{w}_k[n]
d_k[n]+z_j[n]
\end{eqnarray}
where $p_j$ is the allocated power to the $j^{th}$ user. A more detailed compact system formulation
is obtained by stacking the received signals and the noise
components for the set of K selected users as
\begin{eqnarray}
\mathbf{y}[n]=\mathbf{H}\mathbf{W}[n]\mathbf{P}^{\frac{1}{2}}[n]\mathbf{d}[n]+\mathbf{z}[n]
\end{eqnarray}
with $\mathbf{H} = [\mathbf{h}_1,..., \mathbf{h}_K]^T \in\mathbb{C}^{K\times M} $, $\mathbf{W}=[\mathbf{w}_1, ...,\mathbf{w}_K]\in\mathbb{C}^{nt\times M}$ as the
compact channel and precoding matrices. Notice that the transmitted signal $\mathbf{d}\in\mathbb{C}^{K\times 1}$
includes the uncorrelated data symbols $d_k$ for all users with $\mathbb{E}[{|d_k|^2}] = 1$, $\mathbf{P}^{\frac{1}{2}}[n]$
is the power allocation matrix $\mathbf{P}^{\frac{1}{2}}[n]=diag(\sqrt{p_1[n]},\hdots,\sqrt{p_K[n]})$.
It should be noted that CSI and DI are available at the transmitter side. From now on, we assume that the precoding design is performed at each symbol period and accordingly we drop the time index for the ease of notation.
        
 \section{Constructive Interference}
 \label{constructive}

In symbol level precoding (e.g. M-PSK), interference can be constructed in advance in order to push the received symbols further into the correct detection region and, as a consequence it enhances the system performance. Therefore, the interference can
be classified into constructive or destructive based on whether it facilitates or deteriorates the correct detection of the received symbol. For BPSK and QPSK scenarios, a detailed classification of interference is discussed thoroughly in \cite{Christos-1}. In this section, we describe the required conditions to have constructive interference for any M-PSK modulation.
\vspace{-0.1cm}
\subsection{Constructive Interference Definition}

Assuming both DI and CSI are available at the transmitter, the unit-power
 created interference from the $k^{th}$ data stream on $j^{th}$ user can be formulated as:
\vspace{-0.1cm}
\begin{equation}
\psi_{jk}=\frac{\mathbf{h}_{j}\mathbf{w}_k}{\|\mathbf{h}_{j}\|\|\mathbf{w}_k\|}.
\end{equation}
Since the adopted modulations are M-PSK ones, a definition for
constructive interference can be stated as\smallskip

\begin{newtheorem}*{lemma}{\textbf{Lemma}\cite{maha_TSP}}
\begin{lemma}
\label{lemma}
For any M-PSK modulated symbol $d_k$, it is said to receive constructive
interference from another simultaneously transmitted symbol $d_j$ which is
associated with $\mathbf{w}_j$ if and only if the following inequalities hold   
\begin{equation}\nonumber
\label{one}
\angle{d_j}-\frac{\pi}{M}\leq \arctan\Bigg(\frac{\mathcal{I}\{\psi_{jk}d_{k}\}}{\mathcal{R}\{\psi_{jk}d_{k}\}}\Bigg)\leq \angle{d_j}+\frac{\pi}{M},
\end{equation}
\begin{equation}\nonumber
\label{two}
\mathcal{R}\{{d_k}\}.\mathcal{R}\{\psi_{jk}
d_{j}\}>0, \mathcal{I}\{{d_k}\}.\mathcal{I}\{\psi_{jk}d_{j}\}>0.\\
\end{equation}
\end{lemma}
\end{newtheorem}

\begin{newtheorem}*{cor}{\textbf{Corollary}\cite{maha_TSP}}
\begin{cor}
The constructive interference is mutual.
If the symbol $d_j$ constructively interferes with $d_k$, then
the interference from transmitting  the symbol $d_k$ 
is constructive to $d_j$.
\end{cor}

\end{newtheorem}

 \section{Constructive Interference for Power Minimization}
 \label{powmin}
 \subsection{Constructive Interference Power Minimization Precoding with Strict Constellation Transmissions (CIPM) \cite{maha_TSP}}

  From the definition of constructive interference, we should design the constructive interference precoders by granting that the sum of the
precoders and data symbols forces the received
signal to an exact MPSK constellation point namely an exact phase for each user. 
 Therefore, the optimization that
 minimizes the transmit power and grants
 the constructive reception of the transmitted data symbols can be written
 as 
\vspace{-0.2cm}
\begin{eqnarray}
\label{powccm}
\hspace{-0.2cm}\mathbf{w}_k(d_j,\mathbf{H},\boldsymbol\zeta)
\hspace{-0.1cm}&=&\arg\underset{\mathbf{w}_1,\hdots,\mathbf{w}_K}{\min}\quad \|\sum^K_{k=1}\mathbf{w}_kd_k\|^2\\\nonumber
\hspace{-0.1cm}&s.t.&\begin{cases}\mathcal{C}1:\angle(\mathbf{h}_j\sum^K_{k=1}\mathbf{w}_k
d_k)=\angle(d_j),
\forall j\in K\\
\mathcal{C}2:\|\mathbf{h}_j\sum^K_{k=1}\mathbf{w}_kd_k\|^2\geq\sigma^2_n\zeta_j\quad
, \forall j\in K,
\end{cases}
\end{eqnarray}\\
where $\zeta_j$ is the SNR target for the $j^{th}$ user, and ${\boldsymbol\zeta}=[\zeta_1,\hdots,\zeta_K]$ is the vector that contains all the SNR targets.  The set of constraints $\mathcal{C}_1$
gaurantees that the received signal for each user has the correct phase so that the right MPSK symbol $d_j$ can be detected. The solution of eq. (\ref{powccm}) is fully derived in \cite{maha}\cite{maha_TSP}.

\begin{figure}\vspace{0.5cm}\includegraphics[scale=0.27]{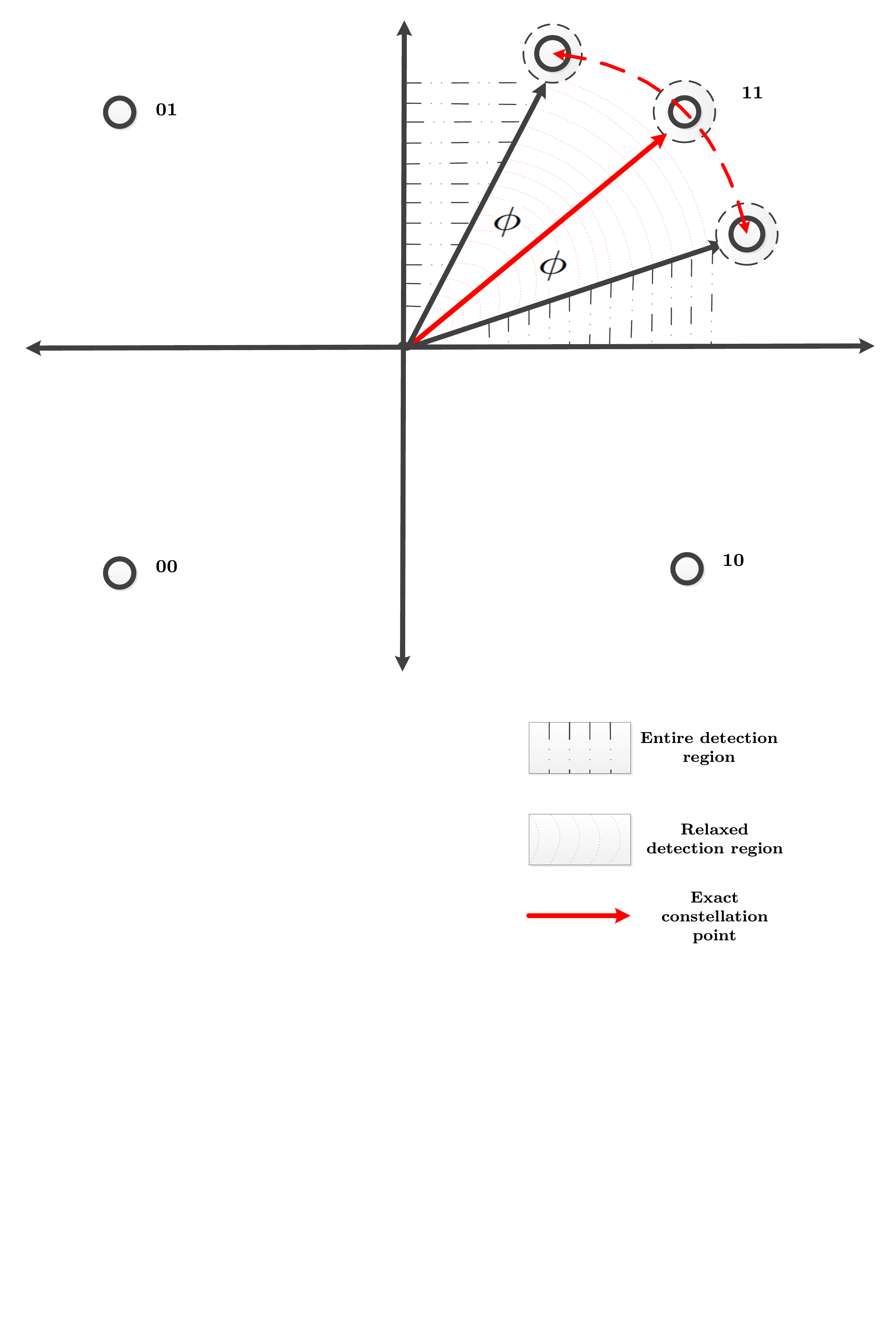}
\vspace{-3.5cm}\caption{\label{qpsk} Constructive Interference Symbol Level Precoding  in  Multiuser MISO Based on Relaxed Detection Region. The phase $\phi$ delimits the relaxed region.  }
\end{figure}

\begin{newtheorem}{thm1}{\textbf{Theorem}} 

\end{newtheorem}

\subsection{Constructive Interference with Relaxed Detection Region (CIPMR)}
To grant a correct M-PSK symbol detection, the  received symbol should lie in the correct
detection region. Fig. (\ref{qpsk}) depicts the detection region of the QPSK symbol $\frac{1+i}{\sqrt{2}}$ which spans the phases $[0^{\circ},90^{\circ}]$.
In the previous section, we design the transmitted symbol to be received with the exact phase
of the target data
symbols except the random deviation resulted from the noise at the receiver; for example the code word was designed so that the symbol $11$ with the phase of $45^{\circ}$. On the other hand, the same symbol can be correctly detected as $11$  with range of phases as long as they lie in the first quadrant and the receiver noise does not push them outside the detection region. Therefore,
it is not necessary to design the transmitted vector $\mathbf{x}$ to have the exact phase
to achieve a correct detection, it can span the set of  $0\leq 45-\phi \leq 45+\phi\leq 90$.
Therefore, more flexibility for the system design can be obtained and more gains
are anticipated.
Since the detection region of symbols span different phases, we can utilize
this property by relaxing the transmitted constellation point to include these angles. The relaxed optimization
can be formulated as
\vspace{-0.4cm}
\hspace{-0.7cm}\begin{eqnarray}\nonumber
\label{CIPMR}
&\hspace{-0.6cm}\mathbf{w}_j(\mathbf{H},\mathbf{d},\mathbf{\zeta},\mathbf{\Phi}_1,\mathbf{\Phi}_2)&\hspace{-0.1cm}=\arg\underset{\mathbf{w}_j}{\min}\quad\|\sum^K_{j=1}\mathbf{w}_jd_j\|^2\\\nonumber
&\hspace{-3.0cm}s.t.&\hspace{-2.7cm}\begin{cases}\mathcal{C}_1:\angle(\underset{\phi^{'}_{j1}}{\underbrace{d_j-\phi_{j1}}})\leq\angle(\mathbf{h}_j\sum^K_{j=1}\mathbf{w}_jd_j)\leq\angle(\underset{\phi^{'}_{j2}}{\underbrace{d_j+\phi_{j2}}}),\forall j\in K\\
\mathcal{C}_2:\|\mathbf{h}_j\sum^K_{j=1}\mathbf{w}_jd_j\|^2\geq\sigma^2\zeta_j\quad
\forall j\in K.\end{cases}
\end{eqnarray}
The problem can be expressed as
\begin{eqnarray}\nonumber
\label{powminr}
&\mathbf{x}&(\mathbf{H},\mathbf{d},\mathbf{\zeta},\mathbf{\Phi}_1,\mathbf{\Phi}_2)\hspace{-0.1cm}=\arg\underset{\mathbf{x}}{\min}\quad\|\mathbf{x}_r\|^2\\\nonumber
&s.t.&\begin{cases}\mathcal{C}_1:\angle(\underset{\phi^{'}_{j1}}{\underbrace{d_j-\phi_{j1}}})\leq\angle(\mathbf{h}_j\mathbf{x})\leq\angle(\underset{\phi^{'}_{j2}}{\underbrace{d_j+\phi_{j2}}}),\forall j\in K\\
\mathcal{C}_2:\|\mathbf{h}_j\mathbf{x}\|^2\geq\sigma^2\zeta_j\quad
\forall j\in K.\end{cases}
\end{eqnarray}
where $\phi_{j1}$ and $\phi_{j2}$ are the phase that received symbol
should be without the noise drifting, $\boldsymbol{\phi}_1$ and $\boldsymbol{\phi}_2$
are the vectors that 
contain all $\phi_{j1}$ and $\phi_{j2}$ respectively. Although this relaxes the phase constraints on the constructive interference design,
it increases the system susceptibility to noise. Therefore, this phase relaxation should
be related to the SNR targets to guarantee certain power saving and SER by  selecting  the allowable values of $\phi_{j1}$ and $\phi_{j2}$. The optimization can be written\footnote{$\pm$ in ($\mathcal{C}_2$-\ref{CIPMR}) indicates that the sign can be positive or negative depending on the value of $\sin\phi$  function. Moreover, $\gtreqless$ is used to indicate that different symbols have different signs (positive or negative) and thus to indicate the correct detection region.}
\begin{eqnarray}
\label{relaxedp}
&\mathbf{x}_r&(\mathbf{H},\mathbf{d},\mathbf{\zeta},\mathbf{\Phi}_1,\mathbf{\Phi}_2)=\arg\underset{\mathbf{x}}{\min}\quad\|\mathbf{x}_r\|^2\\\nonumber
&s.t.&\begin{cases}
\mathcal{C}_1:\mathbf{h}_j\mathbf{x}_r+\mathbf{x}^H_r\mathbf{h}^H_j\gtreqless2\sqrt{\zeta_j}u_j,\forall j\in K \\
\mathcal{C}_2:\mathbf{h}_j\mathbf{x}_r-\mathbf{x}^H_r\mathbf{h}^H_j\gtreqless\pm2i\sqrt{\zeta_j}\sqrt{1-u^2_j},\forall j\in K\\
\mathcal{C}_3:\cos(\phi^{'}_{j2})\leq u_j\leq \cos(\phi^{'}_{j1}), \forall j \in K
\end{cases}
\end{eqnarray}

This optimization has $3K$ constraints that need to be satisfied. The Lagrangian for this problem can be written as 
\begin{eqnarray}\nonumber\label{q}\mathcal{L}(\mathbf{x}_r)&=&\|\mathbf{x}_r\|^2+\sum_j\alpha_j(\mathbf{h}_j\mathbf{x}_r+\mathbf{x}^H_r\mathbf{h}^H_j-2\sqrt{\zeta_j}u_j)\\\nonumber
&+&\sum_j\mu_j(\mathbf{h}_j\mathbf{x}_r-\mathbf{x}^H_r\mathbf{h}^H_j-2i\sqrt{\zeta_j}\sqrt{1-u^2_j})\\\nonumber
&+&\sum_{j}\alpha_j(u_j-\cos(\phi^{'}_{j,1})))+\sum_{j}\gamma_j(u_j-\cos(\phi^{'}_{j,2})))
\end{eqnarray}

\begin{figure*}[t]
\hspace{0.2cm}
\begin{tabular}[t]{c}

\hline
\hline
\end{tabular}
\end{figure*}
 By differentiating
$\mathcal{L}(\mathbf{x}_r)$ with respect to $\mathbf{x}_r^*$ and $u_j$
\begin{eqnarray}\nonumber
\vspace{-0.5cm}\frac{d\mathcal{L}(\mathbf{x}_r,u_j)}{d\mathbf{x}_r}&=&\mathbf{x}+\sum_{j}\alpha_j\mathbf{h}^H_j-\sum_{j}\mu_j\mathbf{h}^H_j,\\
\frac{d\mathcal{L}(\mathbf{x}_r,u_j)}{du_j}&=&-2\sqrt{\zeta_j}+2i\sqrt{\zeta_j}\frac{u_j}{\sqrt{1-u^2_i}}+\lambda_j+\gamma_j.
\end{eqnarray}
By equating $\frac{d\mathcal{L}(\mathbf{x}_r,u_j)}{d\mathbf{x}^*_r}=0$ and $\frac{d\mathcal{L}(\mathbf{x}_r,u_j)}{d\mathbf{u}_j}=0$, we can get the following expressions
\begin{eqnarray}
\label{rw1}
\mathbf{x}_r=\sum_j-\alpha_j\mathbf{h}^H_j+\mu_j\mathbf{h}^H_j
\end{eqnarray}
\begin{eqnarray}
\label{rw2}
u_j=\pm\frac{2\sqrt{\zeta_j}-\lambda_j-\gamma_j}{\sqrt{-4\sqrt{\zeta_j}(\lambda_j+\gamma_j)+\lambda^2_j+2\lambda_j\gamma_j+\gamma^2_j}}
\end{eqnarray}
Substituting (\ref{rw1})-(\ref{rw2}) in the constraints, we have the set of inequalities
(\ref{multicasteq}). It can be noted that the solution of (\ref{powccm}) is a special case of (\ref{relaxedp}) when $\phi_{j1}$ and $\phi_{j2}$ are equal
to zero.\smallskip

\begin{figure*}[t]
\begin{tabular}[t]{c}
\begin{minipage}{18 cm}
 \begin{eqnarray}
\label{multicasteq}
\vspace{-0.5cm}
\begin{array}{cccc}
\label{setoo}
0.5K\|\mathbf{h}_1\|(\sum_k(-\mu_k+\alpha_ki)\|\mathbf{h}_k\|\rho_{1k}&-&\sum_k(-\mu_k+\alpha_ki)\|\mathbf{h}_k\|\rho^{*}_{1k})=\sqrt{\zeta^{}_{1}}\sqrt{1-\frac{(2\sqrt{\zeta_1}-\lambda_1-\gamma_1)^2}{{-4\sqrt{\zeta_1}(\lambda_1+\gamma_1)+\lambda^2_1+2\lambda_1\gamma_1+\gamma^2_1}}}\\
0.5K\|\mathbf{h}_1\|(\sum_k(-\mu_ki-\alpha_k)\|\mathbf{h}_k\|\rho_{1k}&+&\sum_k(-\mu_ki-\alpha_k)\|\mathbf{h}_k\|\rho^{*}_{1k})=\sqrt{\zeta^{}_{1}}\frac{2\sqrt{\zeta_1}-\lambda_1-\gamma_1}{\sqrt{-4\sqrt{\zeta_1}(\lambda_1+\gamma_1)+\lambda^2_1+2\lambda_1\gamma_1+\gamma^2_1}}\\
\quad&\vdots&\\
0.5K\|\mathbf{h}_K\|(\sum_k(-\mu_k+\alpha_ki)\|\mathbf{h}_k\|\rho_{Kk}&-&\sum_k(-\mu_k+\alpha_ki)\|\mathbf{h}_k\|\rho^{*}_{Kk})=\sqrt{\zeta_{K}}\sqrt{1-\frac{(2\sqrt{\zeta_K}-\lambda_K-\gamma_K)^2}{{-4\sqrt{\zeta_K}(\lambda_K+\gamma_K)+\lambda^2_K+2\lambda_K\gamma_K+\gamma^2_K}}}\\
0.5K\|\mathbf{h}_K\|(\sum_k(-\mu_ki-\alpha_k)\|\mathbf{h}_k\|\rho_{Kk}&+&\sum_k(-\mu_ki-\alpha_k)\|\mathbf{h}_k\|\rho^{*}_{Kk})=\sqrt{\zeta_{K}}\frac{2\sqrt{\zeta_K}-\lambda_K-\gamma_K}{\sqrt{-4\sqrt{\zeta_K}(\lambda_K+\gamma_K)+\lambda^2_K+2\lambda_K\gamma_K+\gamma^2_k}}\\
{2\sqrt{\zeta_1}-\lambda_1-\gamma_1}&\leq& {\sqrt{-4\sqrt{\zeta_1}(\lambda_1+\gamma_1)+\lambda^2_1+2\lambda_1\gamma_1+\gamma^2_1}} \cos(\phi^{'}_{11})\\
{2\sqrt{\zeta_1}-\lambda_1-\gamma_1}&\geq& {\sqrt{-4\sqrt{\zeta_1}(\lambda_1+\gamma_1)+\lambda^2_1+2\lambda_1\gamma_1+\gamma^2_1}}\cos(\phi^{'}_{12})\\
\quad&\vdots&\\
{2\sqrt{\zeta_K}-\lambda_K-\gamma_K}&\leq& {\sqrt{-4\sqrt{\zeta_K}(\lambda_K+\gamma_K)+\lambda^2_K+2\lambda_K\gamma_K+\gamma^2_K}}\cos(\phi^{'}_{K1})\\
{2\sqrt{\zeta_K}-\lambda_K-\gamma_K}&\geq& {\sqrt{-4\sqrt{\zeta_K}(\lambda_K+\gamma_K)+\lambda^2_K+2\lambda_K\gamma_K+\gamma^2_K}}\cos(\phi^{'}_{K2})
\end{array}
\end{eqnarray}
\normalsize
\end{minipage}\\
\vspace{-0.3cm}\\
\hline
\hline
\end{tabular}
\normalsize
\end{figure*}
\subsubsection{Simple solution}
Another simple solution can be found for the scenario of $\phi_{j1}=\phi_1,\forall j\in K$, $\phi_{j2}=\phi_2,\forall j\in K$ and $\phi_1=\phi_2$ by searching all the phases that lie in the relaxed region. The linear search is performed on the value of $\phi_i$ which is varied from $\angle d_j-\phi$ to $\angle d_j+\phi$ to achieve the minimum power consumption. For each value $\phi_i\in [\angle d_j-\phi,\angle d_j+\phi]$, we solve the following optimization

\begin{eqnarray}
\label{relaxedp}
&\mathbf{x}_r&(\mathbf{H},\mathbf{d},\mathbf{\zeta},\phi_i)=\arg\underset{\mathbf{x}}{\min}\quad\|\mathbf{x}_r\|^2\\\nonumber
&s.t.&\begin{cases}
\mathcal{C}_1:\mathbf{h}_j\mathbf{x}_r+\mathbf{x}^H_r\mathbf{h}^H_j\gtreqless2\sqrt{\zeta_j}\cos(\angle d_j+\phi),\forall j\in K \\
\mathcal{C}_2:\mathbf{h}_j\mathbf{x}_r-\mathbf{x}^H_r\mathbf{h}^H_j\gtreqless2i\sqrt{\zeta_j}\sin(\angle d_j+\phi),\forall j\in K,\\
\end{cases}
\end{eqnarray}
to find the phase that has the minimum power consumption
\begin{eqnarray}
\phi^*=\underset{\phi_i}{\arg}{\min}f(\phi_i).
\end{eqnarray}
$f(\phi_i)$ is a function that maps the minimum power with its respective phase.
The flexible constellation transmissions provide more freedom to find the proposed CI precoding that requires 
less power to achieve the target SNR. On the other hand, this transmit power reduction comes at the expense of increasing the probability of
 symbol error rate (SER) at the receivers due to the expected noise deviation of the received symbols from their exact point of detection. The trade off between the power saving and SER is studied thoroughly from the system energy efficiency perspective in \cite{maha_TWC}.

\subsection{Constructive Interference Power Minimization Bounds}

In order to assess the performance of the proposed algorithm, we mention
two theoretical upper bound as follows

\subsubsection{Genie aided upper bound}
This bound occurs when all multiuser transmissions are constructively
interfering by nature and without the need to optimize the output vector. If we assume $\mathbf{W}=\mathbf{H}^{'}$, where $\mathbf{H}^{'}=[\frac{\mathbf{h}^H_1}{\|\mathbf{h}_1\|}, \hdots,\frac{\mathbf{h}^H_K}{\|\mathbf{h}_K\|}]$. By exploiting singular value decomposition (SVD) of $\mathbf{H}$. $\mathbf{V}^{'}$ is the power scaled of $\mathbf{V}$
to normalize each column in $\mathbf{W}$ to unity. The received signal can be as
\begin{eqnarray}
\label{svd}
\vspace{-0.2cm}
\mathbf{y}&=&\mathbf{H}\mathbf{W}\mathbf{d}={{\mathbf{S}\mathbf{V}\mathbf{D}\mathbf{D}^H\mathbf{V}^{'}}}{{\mathbf{S}^H}}\mathbf{P}^{1/2}\mathbf{d}.
\end{eqnarray} 
If we denote $\mathbf{G}=\mathbf{SVD}\mathbf{D}^H$ and $\mathbf{B}=\mathbf{S}^H$. Utilizing the reformulation of $\mathbf{y}$ in
(\ref{svd}), the received signal can be written as 
\vspace{-0.05cm}
\begin{eqnarray}
\label{rot}
y_j=\|\mathbf{g}_j\|\sum^K_{k=1}\sqrt{p_k}\xi_{jk}d_k,
\end{eqnarray}
where $\mathbf{g}_j$ is the $j^{th}$ row of the matrix $\mathbf{G}$, $\xi_{jk}=\frac{\mathbf{g}_j\mathbf{b}_k}{\|\mathbf{g}_j\|}$. 
  The minimum transmit power for a system that exploits the constructive interference
on symbol basis can be found by the following theoretical bound
\begin{thm1}
The genie-aided minimum transmit power in the downlink of multiuser MISO   system can be found by solving the following optimization
\begin{eqnarray}
\label{pr}
\vspace{-0.1cm}
&{P}_{min}&=\arg\underset{p_1,\hdots,p_K}{\min}\quad\sum^K_{k=1}p_k\\\nonumber
&s.t.&\|\mathbf{g}_k\|^2(|\xi_{kk}|^2{p_k}+\sum^K_{j=1,j\neq k}{p_j}|\xi_{kj}|^2)\geq{\zeta_k},\forall
k\in K.
\end{eqnarray}

\end{thm1}


\subsubsection{Optimal Multicast} 
Based on theorem (2), a theoretical upperbound can be characterized. This bound occurs if we drop the phase alignment
constraint $\mathcal{C}_1$. The intuition of
using this technique is the complete correlation among the information that needs to be communicated (i.e. same symbol for all users). The optimal input covariance for power minimization in multicast system can
be found as a solution of the following optimization
\vspace{-0.05cm} 
\begin{eqnarray}
\label{powm1}
&\underset{\mathbf{Q}:\mathbf{Q}\succeq 0}{\min}&\quad tr(\mathbf{Q})\quad s.t.\quad\mathbf{h}_j\mathbf{Q}\mathbf{h}^H_j\geq\zeta_j\quad,\forall j\in K.
\end{eqnarray}
This problem is thoroughly solved in \cite{multicast}. 
\vspace{-0.2cm}
\section{Numerical results}\label{sim}

\vspace{-0.15cm}
In order to assess the performance of the proposed transmissions schemes, Monte-Carlo simulations of the different algorithms have been conducted to
study the performance of the proposed techniques and compare to the state
of the art techniques. The adopted channel model is assumed
to be $\mathbf{h}_k\sim\mathcal{CN}(0,\sigma^2)$. For the sake of comparison, the system energy efficiency is used and can be defined as 
\begin{eqnarray}
\label{metric}
\eta= \frac{\sum^K_{j=1} \bar{R}_j\Big(SER_j(\zeta_j,\phi_j)\Big)}{\|\mathbf{x}(\mathbf{H},\mathbf{d},\boldsymbol\zeta,\boldsymbol\Phi)\|^2}
\end{eqnarray}
where $\bar{R}_j\approx R_j\times \big(1-SER (\zeta_j,\phi_j)\big)$, ${R}_j$ is the rate associated with MPSK modulation, $\bar{R}_j$ is the effective rate for $j^{th}$ user or goodput that takes into the account the symbol error rate $SER_j$ for the $j^{th}$ user.

The comparison among optimal multicast, CIPM and CIPMR is illustrated in this figure while the assumed scenario is $M=3$, $K=2$, we study the performance of CIPMR at $\phi=\frac{\pi}{5}$ and $\frac{\pi}{8}$. It can concluded that the power consumption gap between the optimal multicast and CIPM is fixed for all target rates. This relation holds also for the gap between the CIPMR and CIPM. Moreover, it can be concluded that the CIPMR outperforms CIPM by achieving less power at all target SINR values. Moreover, CIPMR
 shows a better performance at $\phi=\frac{\pi}{5}$  than $\phi=\frac{\pi}{8}$.
 
 \begin{figure}[h]
\begin{center}
\vspace{-0.1cm}
\includegraphics[scale=0.59]{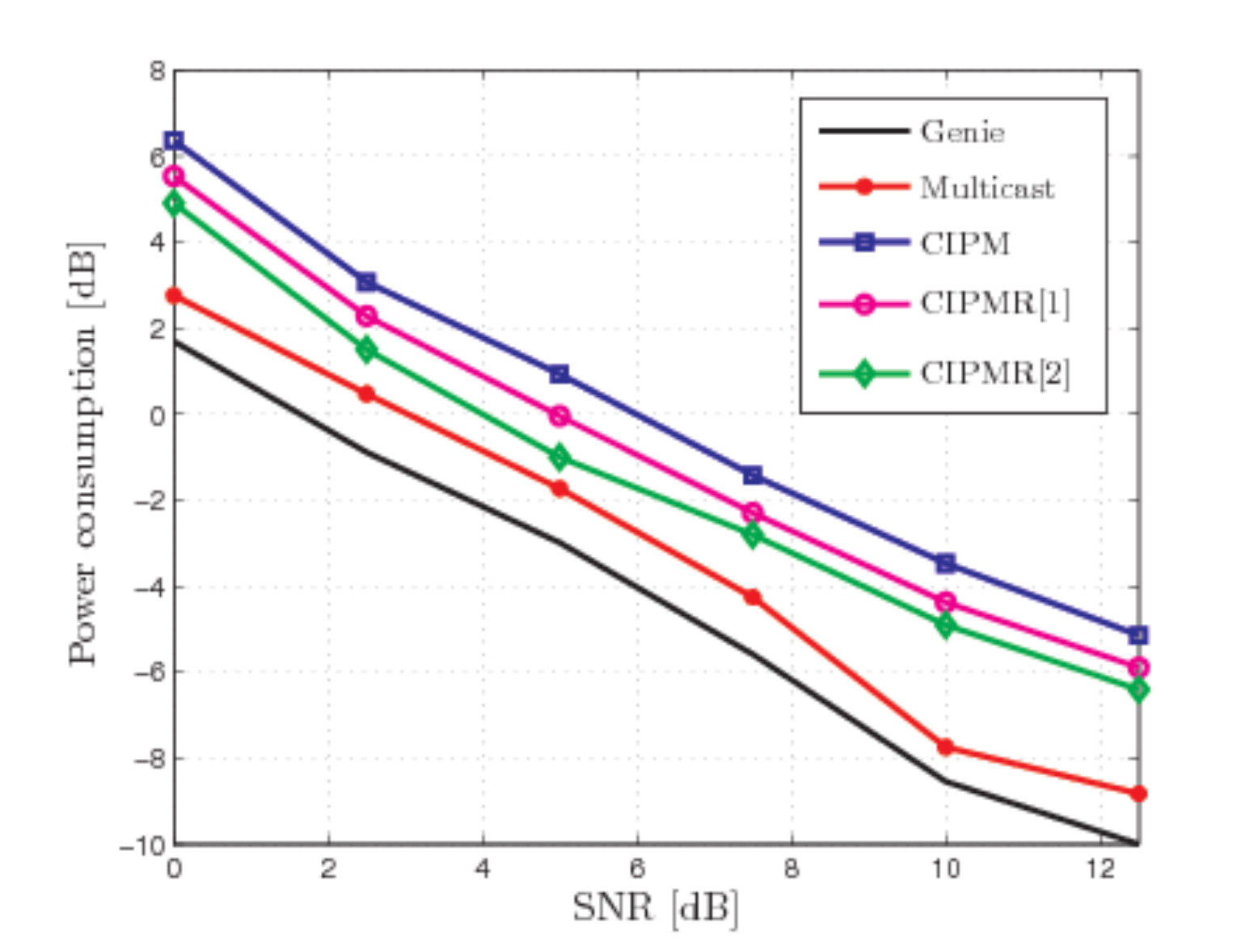}
\vspace{-0.1cm}\caption{\label{powerconsumption}Power consumption vs channel SNR. CIPMR[1] denotes the scenario of $\phi=\frac{\pi}{5}$ and CIPMR[2] denotes the scenario of $\phi=\frac{\pi}{8}$.}
\end{center}
\end{figure}

 \begin{figure}[h]
 \vspace{-0.1cm}
\begin{center}
\vspace{-0.2cm}
 \includegraphics[scale=0.57]{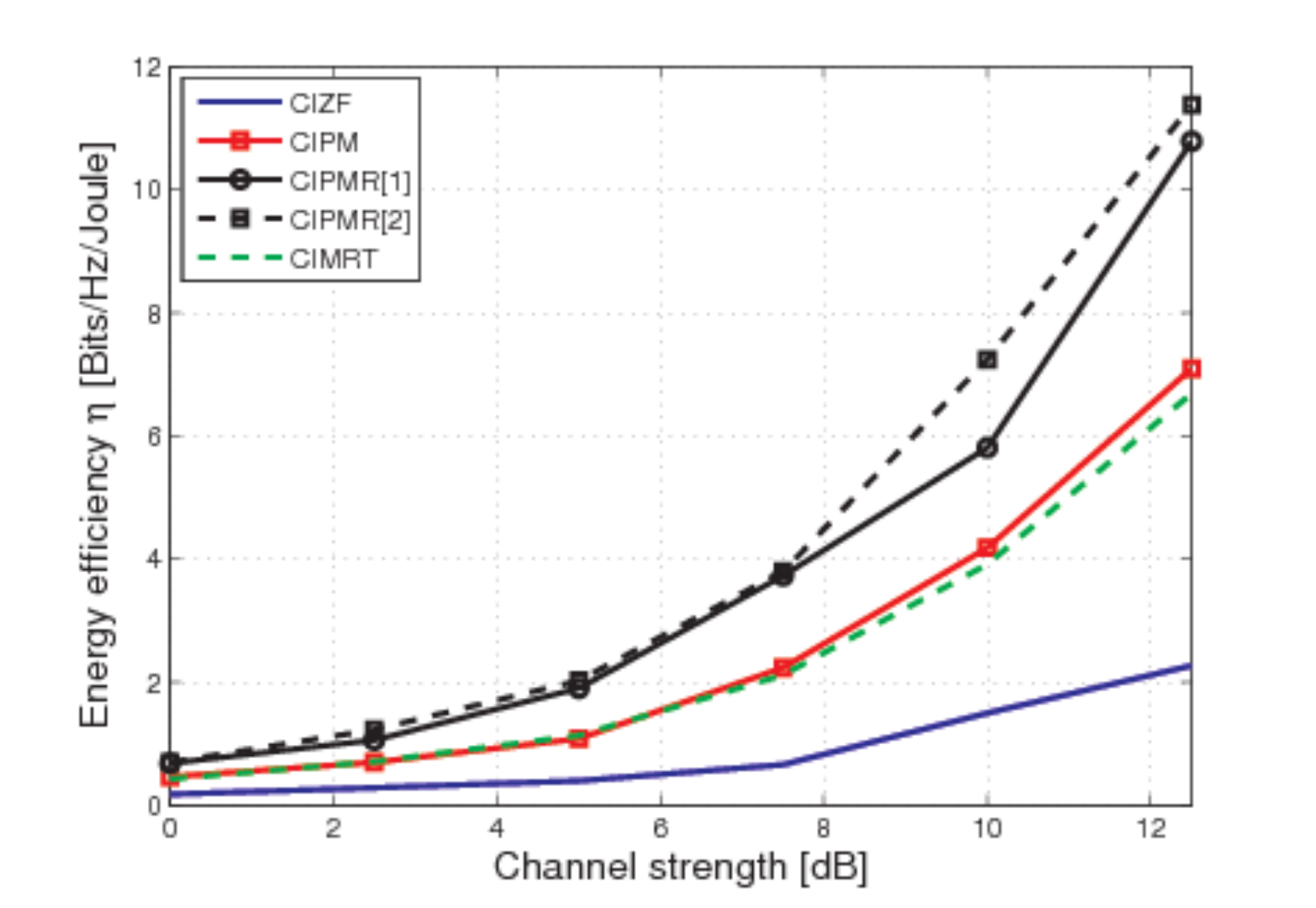}
\vspace{-0.2cm}\caption{\label{eevs}Rate per user vs. channel strength. CIPMR[1] denotes $\phi=\frac{\pi}{8}$ and CIPMR[2] denotes the scenario of $\frac{\pi}{5}$. $M=3$, $K=2$, $\zeta=4.7121 dB$, QPSK.}
 \end{center}
 \end{figure}
\begin{center}
\begin{table}
\hspace{-0.02cm}\begin{tabular}{|p{1.3cm}|p{5.2cm}|p{1.0cm}|}
\hline
Acronym&Technique&equation\\
\hline
CIZF&Constructive Interference Zero Forcing&\cite{Christos}\\
\hline
CIMRT&Constructive Interference Maximum Ratio Transmissions&\cite{maha}
\\
\hline
CIPM& Constructive Interference- Power Minimization&\ref{powccm},\cite{maha},\cite{maha_TSP}\\
\hline
CIPMR& Constructive interference power minimization with relaxed constellation& \ref{CIPMR}\\
\hline
Multicast&Optimal Multicast &\ref{powm1}\\
\hline
\end{tabular}
\vspace{0.2cm}
\caption{Summary of the proposed algorithms, their related acronyms, and
their related equations and algorithms}
\end{table}
\end{center}
In Fig. (\ref{eevs}), we depicted the performance of the proposed techniques from energy efficiency
perspective with the channel strength. CIZF shows inferior performance in comparison with all
depicted techniques. It has already been proven that CIZF 
outperforms the conventional techniques like minimum mean square error (MMSE)
beamforming and zero forcing beamforming (ZFB) \cite{Christos}. In comparison with other depicted techniques, it can be concluded that
the proposed constructive interference CIPM and CIPMR have better energy efficiency in comparison with CIZF.
This can be explained by the channel inversion step in CIZF which wastes energy in
decoupling the effective users' channels and before exploiting the
interference among the multiuser streams. Furthermore, it can be deduced that
CIMRT has a very close performance to CIPM especially at high targets. CIMRT
outperforms CIZF at expense of complexity.

Most importantly, it can be noted that CIPMR achieves higher energy efficiency than CIPM at different average channel power values.  It can be concluded that despite increasing the phase margin decreases the effective rate, it decreases the power consumption. Therefore, the energy efficiency is function of the phase margin and can optimized to find the optimal phase margin at certain SNR target and average channel power. 

\vspace{-0.2cm}
\section{conclusions}
In this paper, we study the possibility of exploiting the interference among the multiuser transmissions in the downlink of multiple antennas base stations. We propose a symbol based precoding that uses M-PSK modulation to exploit the interference among the multiuser transmissions. Particularly, we utilize the concept that the detection region of an M-PSK symbol spans the range of phases, this enables us to relax the system design to achieve more power savings.  This can be implemented by allowing the precoder to select the optimal phase for each user symbols that can achieve the minimum power without being erroneously detected at the receiver. From the simulation, it can be noticed that the relaxed system designs achieve less power consumption and higher energy efficiency than the strict constellation points.

\end{document}